# CFPB Consumer Complaints Analysis Using Hadoop

Dhwani Vaishnav, Manimozhi Neethinayagam, Akanksha S Khaire, Mansi Vivekanand Dhoke, Jongwook Woo

Department of Information Systems, California State University Los Angeles
{dvaishn2, mneethi, akhaire3, mdhoke, jwoo5}@calstatela.edu

**Abstract**

Consumer complaints are a crucial source of information for companies, policymakers, and consumers alike. They provide insight into the problems faced by consumers and help identify areas for improvement in products, services, and regulatory frameworks.

This paper aims to analyze Consumer Complaints Dataset provided by Consumer Financial Protection Bureau (CFPB) and provide insights into the nature and patterns of consumer complaints in the USA. We begin by describing the dataset and its features, including the types of complaints, companies involved, and geographic distribution. We then conduct exploratory data analysis to identify trends and patterns in the data, such as the most common types of complaints, the companies with the highest number of complaints, and the states with the most complaints. We have also performed descriptive and inferential statistics to test hypotheses and draw conclusions about the data. We have investigated whether there are significant differences in the types of complaints or companies involved based on geographic location. Overall, our analysis provides valuable insights into the nature of consumer complaints in the USA and helps stakeholders make informed decisions to improve the consumer experience.

## 1. Introduction

The aim of this paper is to analyze consumer complaints in the USA using Big Data platform, Hadoop, and Hive, specifically focusing on finance-related issues faced by consumers in financial institutions. We utilized the Consumer Financial Protection Bureau Dataset provided by the U.S. government agency dedicated to consumer protection. The dataset was used because it provides data on finance-related issues faced by consumers in the USA and is made available by a U.S. government agency dedicated to consumer protection in banks and financial institutions. By analyzing this dataset, we can gain valuable insights into consumer complaints and identify trends, patterns, and relationships that can help improve consumer protection and financial services. Additionally, the dataset is quite large, at 3.6 GB, making it suitable for use in big data platform. Our analysis includes analysis of financial organizations & their services receiving high concerns, year-on-year complaints growth rate analysis and overall analysis of the sentiment of the consumers. As California residents, we have also focused on complaints statistics for the state of California. The results of this analysis provide valuable information on consumer complaints trends and issues, which can be used to improve the customer experience in the financial industry. The Consumer Financial Protection Bureau (CFPB) dataset used in this analysis is publicly accessible and intended for public use.

## 2. Related Work

In recent years, there has been a growing interest in analyzing the Consumer Financial Protection Bureau (CFPB) dataset to better understand consumer complaints in the financial industry. One such related work is a study by Chen and Zhang [1] that focuses on the impact of the Dodd-Frank Wall Street Reform and Consumer Protection Act on consumer financial outcomes. They utilized the CFPB database to examine consumer complaints before and after the implementation of the Act, finding that the Act had a positive impact on consumer financial outcomes by reducing complaints related to predatory lending, mortgage servicing, and debt collection. Another example is a study by Rupesh et al. [2] that analyzes the CFPB data to identify patterns and trends in mortgage, debt collection, and credit reporting complaints. Their analysis, performed using Tableau Software and IBM Watson Analytics, found that complaints were constantly increasing between 2011 and June 2016, with the largest number of complaints in California involving credit card disputes. Another example of a dataset analysis using the Consumer Financial Protection Bureau (CFPB) dataset was conducted by P. Van Leuven, K. Miller, and S. Spitzer [3]. In their research, they aimed to identify patterns and relationships in consumer complaints regarding mortgage loans, and to evaluate the effectiveness of the CFPB's complaint handling process. Their insights provide an analysis of complaints received in 2021, categorized by product and service, geographic region, and special population such as servicemembers and older consumers. They also evaluated the timeline and accuracy of the CFPB's complaint handling process by comparing the results of their analysis to the CFPB's official reports. The study found that the most common issues reported by consumers were related to loan modification and servicing. This study provides valuable insights into consumer complaints related to mortgage loans and the effectiveness of the CFPB's complaint handling process.

In contrast to these related works, our analysis focused on all U.S. business complaint data from 2021 to 2023, plotting year-on-year growth in complaints, and filtering CFPB complaint data based on the highest-receiving financial organizations, as well as only including complaints from California businesses. We also analyzed overall sentiment in complaints and used Ngram Text Processing in the consumer narrative section to obtain a set of frequently used words for further analysis. We visualized the data using Tableau Software and, for sentiment analysis, Excel Power Map. Overall, our analysis provides a unique perspective on the CFPB dataset and offers valuable insights into consumer complaints and trends in the financial industry.

## 3. Specifications

The Consumer Complaint Database is a collection of complaints about consumer financial products and services that the Consumer Financial Protection Bureau (CFPB) receives from consumers. The dataset contains detailed information about each complaint, including the date of submission, the consumer's zip code, the type of financial product or service being complained about, and the nature of the complaint. The dataset is continuously updated and as of the date we downloaded, it's of the size 3.6 GB. It contains complaints data from 2011 to March 2023.

Below Table 1 shows files and size of the files from dataset.

*Table 1 Data Specification*

| Data Set Size | 3.6 GB |
|---|---|
| Number for files | 1 |
| Content Format | JSON |

The Table 2 below shows the specification for Oracle cluster we are using and Hadoop specification for our project.

*Table 2 H/W Specification*

| Number of nodes | 5 (2 master nodes, 3 worker nodes) |
|---|---|
| CPU speed | 1995.312 MHz |
| Storage | 390 GB |

## 4. Implementation Flowchart

Initially, the raw dataset, which comprises the detail of consumer complaints from CFPB platform, was downloaded from data.gov.

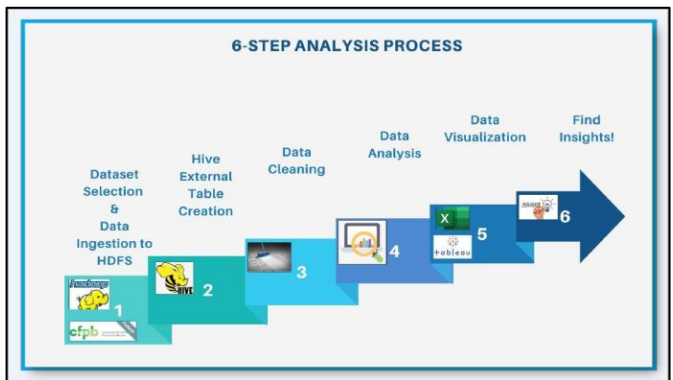

*Figure 1 Architecture Flow Chart*

The whole process of data manipulation is shown in the *Figure 1 Architecture Flow Chart*.

The dataset is available in two different formats – CSV & JSON. We downloaded the dataset in JSON format and uploaded it to the Hadoop File System. After that, HiveQL is used as querying language to create the tables' schema, clean data, create a summary table and export the results. Once the output file has been downloaded in CSV/TSV format, we used Excel's 3D map and Tableau to obtain the visualizations.

## 5. Data Cleaning

Raw files were uploaded and stored in HDFS and then loaded into tables using Beeline Client. The dataset doesn't contain much NULL or missing values and hence it doesn't need to be thoroughly clean. However, it contains the information regarding the complaints registered by consumers in narration and therefore contains many special characters. Additionally, since this dataset is all about financial services, there are confidential and secure details of the customers present in the dataset which is not made public and enclosed with 'XXXX' wherever required. For handling the special characters and replacing the 'XXXX' from the 'customer narrative' field, data cleaning was conducted using regular expressions.

## 6. Analysis and Visualization

After data cleaning and preparation for further analysis, files were extracted into BI tools, Tableau and Excel. We used different interactive maps to show statistics based on total complaints, the companies receiving those complaints and the services having problems according to the complaints. We have carried out two different types of sentiment analysis for the period of 2021 to 2023 to know the overall polarity of the complaints for each state of United States.

### 6.1 3D Map in Excel

The first visualization Figure 2 Sentiment Analysis, a 3D map, was made in Excel and it is an animated map with a timeline for one year, April 2022 to April 2023. This visualization uses a bubble chart to represent each state on a map, with the size of the bubble indicating the sentiment count (i.e., number of complaints) in that state. The layers with different colors of the bubbles indicates the sentiment (positive, negative, or neutral). The map is arranged by state and the time element is represented by months of date received, allowing us to analyze sentiment trends over time. This visualization helps to identify states with the most complaints for a particular sentiment, and using the 3D map feature, it allows for interactive viewing and analysis from different angles. Most states have more negative sentiment values than positive, this indicates that consumers in these states have experienced more negative outcomes than positives. However, for the states like Texas & California, the distribution is more towards positives, which is an interesting finding! This visualization is in a video format, by playing the video, it is clear to see that the bars grow faster after September 2022.

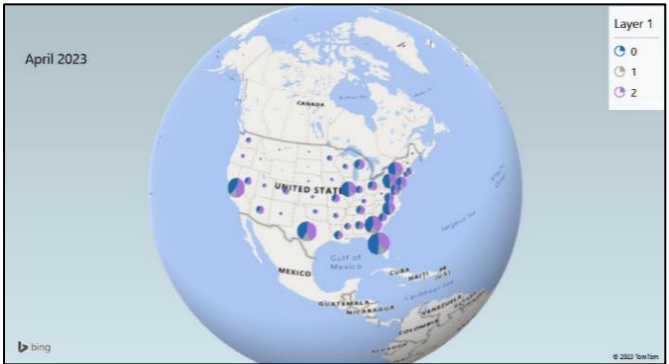

Figure 2 Sentiment Analysis

## 6.2 Tableau

The dashboard in the Figure 3 Overall Complaints Statistics displays the overall statistics of complaints registered in USA. Using HIVE queries, we found out for which company consumers registered maximum complaints and our analysis revealed EQUIFAX, INC has the highest complaints filed against it. We used bubble chart to show the issues faced by the consumers. The product Credit Reporting has received most of complaints from consumers which is 80.19%. There are several mediums available for registering the complaints, however approx. 86% of complaints were received through CFPB website.

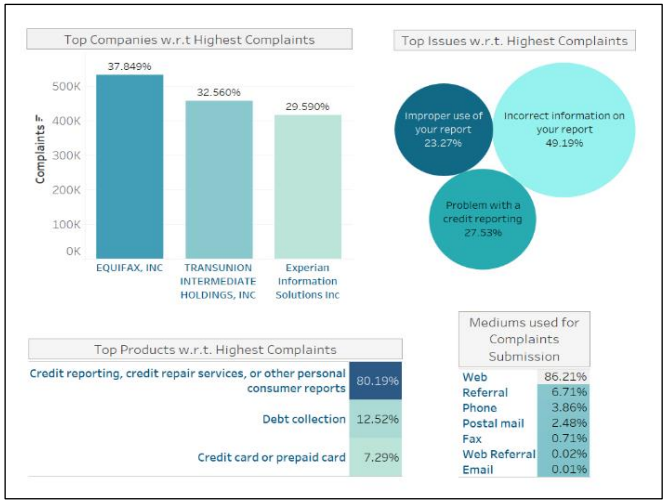

*Figure 3 Overall Complaints Statistics*

The dashboard in the next Figure 4 Year on year Complaints Statistics is created in Tableau. It contains a tree map to display the number of complaints received per quarter from Q1 2021 to Q1 2023. The % difference in complaints count compared to the previous quarter allows us to see how the number of complaints is changing over time. For example, if the % difference is positive, this would indicate an increase in the number of complaints compared to the previous quarter, while a negative % difference would indicate a decrease. This would provide a more nuanced understanding of the data, allowing us to see the rate of change over time. The dashboard also contains the area chart to represent gradual increase in the complaints over the quarters from 117K to 289K.

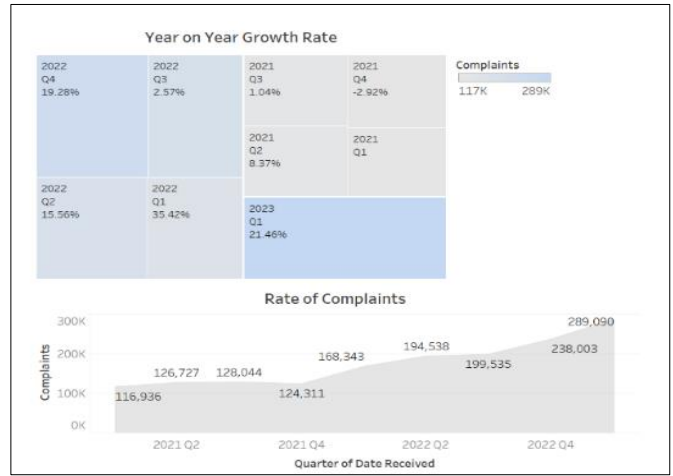

*Figure 4 Year on year Complaints Statistics*

After conducting a tempo-spatial analysis using HIVE queries, we were able to determine which state had the highest volume of customer complaints. According to our findings shown below in Figure 5 State wise distribution of complaints, FLORIDA had the largest number of complaints throughout the United States, with TEXAS coming in second and CALIFORNIA taking third place. In contrast, WYOMING had the lowest number of complaints recorded nationwide.[1]

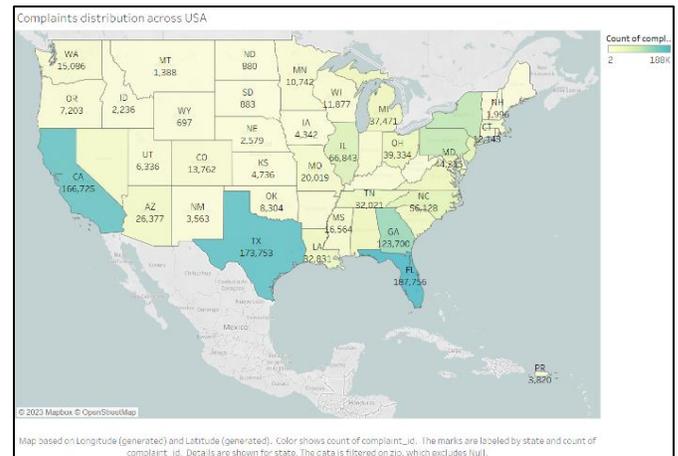

*Figure 5 State wise distribution of complaints*

---

[1] The minor outlying islands are not considered in the list of the states

Despite Florida having the highest number of consumer complaints, our curiosity as California residents led us to analyze our own state. As shown in Figure 6 California Complaints Statistics, our analysis revealed that customers of California registered 166,725 complaints. We conducted HIVE queries to understand which company has the maximum complaints, and found that EQUIFAX, INC has received maximum complaints in California too. We used a Heat map to show this finding. With this information we analyzed the data further to understand for which product there was so many complaints and found an interesting result that 'Credit Reporting' has received 80.59% of complaints, showed in pie chart which is almost 8 times higher than other products. Additionally, for the actions taken by the company for the registered complaints, we also identified that EQUIFAX has closed 91.68% of their complaints with explanation[2] and 0.2% of complaints with non-monetary relief.

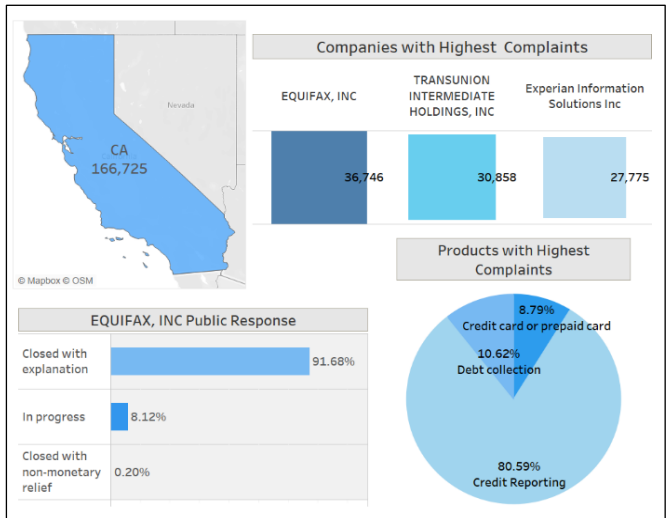

*Figure 6 California Complaints Statistics*

Through our analysis we found EQUIFAX, INC received the maximum number of complaints throughout the USA, so we performed NGRAM TEST PROCESSING on Customer Narrative Section to understand why there are a lot of complaints. We used 4-word count NGRAM, because through bigram and trigram we were not able to get any meaningful insights as there is lot confidential details hidden with "XXXX". Our analysis revealed "*Victim of Identity Theft*" has appeared frequently in the list. EQUIFAX INC needs to address this issue to make their customers satisfied. The result is shown below in *Figure 7 NGram Snippet*.

*Figure 7 NGram Snippet*

## 7. Conclusion

In conclusion, the above analysis highlights the significant impact that EQUIFAX, INC. has had on its customers. The large number of complaints registered by customers, particularly in Florida, Texas, and California, indicates a widespread problem with the company's credit reporting product. The fact that identity theft was the primary reason for customer complaints underscores the importance of safeguarding personal information in today's digital age. Moreover, the negative sentiment expressed in the General Sentimental Analysis suggests that customers have had overwhelmingly negative experiences with the company. Finally, the high proportion of complaints received through the website highlights the need for effective online customer support channels. Overall, this analysis emphasizes the need for EQUIFAX, INC. and other companies to prioritize their customers' needs and concerns to ensure better customer satisfaction.

---

[2] For the complaints 'Closed with explanation' for the 'Public Response' by EQUIFAX, we don't have any data on which if the complaints were closed with monitory/non-monitory benefits.